\documentclass{PoS}

\title{Gluon mass at finite temperature in Landau gauge}

\ShortTitle{Gluon mass at finite temperature in Landau gauge}

\author{\speaker{Pedro Bicudo}\\
         CFTP, Instituto Superior T\'{e}cnico, CFTP, Universidade de Lisboa, 1049-001, Lisboa, Portugal\\
        E-mail: \email{bicudo@tecnico.ulisboa.pt}
        }

\author{Orlando Oliveira\\
       CFC, Departamento de F\'{\i}sica, Faculdade de Ci\^encias e Tecnologia, Universidade de Coimbra, 3004-516 Coimbra, Portugal\\
 E-mail: \email{orlando@fis.uc.pt}
       }
       
\author{Paulo Silva\\
       CFC, Departamento de F\'{\i}sica, Faculdade de Ci\^encias e Tecnologia, Universidade de Coimbra, 3004-516 Coimbra, Portugal\\
 E-mail: \email{psilva@teor.fis.uc.pt}
       }
       
\author{Nuno Cardoso\\
         CFTP, Instituto Superior T\'{e}cnico, CFTP, Universidade de Lisboa, 1049-001, Lisboa, Portugal\\
 E-mail: \email{nunocardoso@cftp.ist.utl.pt}
        }

\abstract{Using lattice results for the Landau gauge gluon propagator at finite temperature, we investigate its interpretation as a massive type bosonic propagator. In particular, we estimate a gluon mass from Yukawa-like fits to the lattice data and study its temperature dependence. }

\FullConference{31st International Symposium on Lattice Field Theory - LATTICE 2013\\
		July 29 - August 3, 2013\\
		Mainz, Germany}

\begin{document}

\section{Introduction}

Lattice QCD not only is quite important to compare QCD with experiment, but also is ideal to test theories, approximations  and models. Here we address pure gauge theory in the phase transition region.

At $T=0$ pure gauge SU(3) QCD exhibits color screening and flux tubes
\cite{Cornwall:1981zr,Oliveira:2010xc,Cardoso:2013lla},
while at large $T$ Debye screening occurs
\cite{Doring:2007uh}.
At $T=T_c \sim 270$ MeV, there is evidence of a finite a gluon mass scale in the $\pi$ and $K$ multiplicities in heavy ions
\cite{Bicudo:2012wt}.

Here we complement the outstanding study 
\cite{Heller:1997nqa}
of the gluon masses in SU(2) for $2 T_c < T < 15000 T_c$.
We study  
\cite{Silva:2013maa}
the finite temperature range $T < 2 T_c$ in pure gauge SU(3).

The study of the gluon propagator and gluon mass require gauge fixing, and we resort to Landau gauge fining.

\begin{figure} \begin{center}
\includegraphics[width=0.6\textwidth]{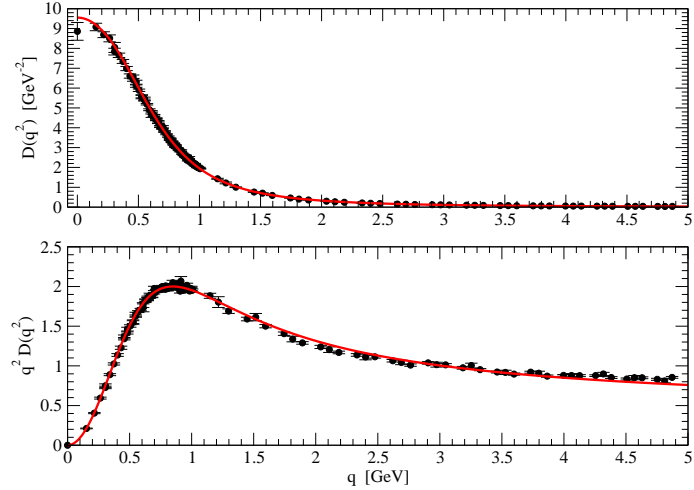}
\end{center}
\caption{Landau gauge gluon propagator at $\beta=6.0$, renormalized at $\mu=3$ GeV, combining $31⁴, \, 48⁴, \, 64^4, \, 80^4$ data.}
\label{fig1}
\end{figure}

\section{Gluon propagator with Landau gauge fixing at T=0}

On the lattice, the Landau gauge fixing is applied to a configuration $U_{\mu
}(\mathbf x)$ by maximizing the function,
\begin{equation}
F_{U}[g] =C_{F}\sum_{\mathbf x,\mu} \mbox{Re} \{\mbox{Tr} [ g(\mathbf x)U_{\mu
}(\mathbf x)g^{\dagger }(\mathbf x+\hat{\mu})]\} \label{f}c\ ,
\end{equation}
where $ g(\mathbf x) $ is a gauge transformation. The maximum leads to, 
\begin{equation}
\partial_\mu A^a_\mu=0  \ .
\end{equation}
We apply a (Fourier accelerated) Steepest Descent method. We have tested
this method both in CPU's and GPU's
\cite{Cardoso:2012pv,Schrock:2012fj}.

We compute the $D(p^2)$, shown  in Fig. \ref{fig1}, with pure gauge lattice simulations, utilizing the Wilson action for pure gluons, and the expectation value, 
\begin{equation}
 \langle A^a_\mu (p) A^b_\nu (p) \rangle = V \, \delta (p - k) \, \delta^{ab} \left( \delta_{\mu\nu} - \frac{p_\mu p_\nu}{p^2} \right) \, D(p^2)
\end{equation}
We utilize sufficiently large volumes $V= {L_s}^3$, since a larger volume implies we can reach smaller infrared (IR) momenta for the computation of  $D(p^2)$. We also use small lattice spacing, to reduce the 
$\mathcal{O}(a^2)$ corrections effects, relevant both in the IR and medium range momenta
\cite{Oliveira:2012eh}.

\begin{table}
\begin{center}
\begin{tabular}{cccccc}
\hline
Temp. (MeV) &	$\beta$ & $L_s$ &  $L_t$ & a [fm] & 1/a (GeV) \\
\hline
121 &   6.0000 & 64 & 16 & 	0.1016 &  	1.9426 \\
162 &   6.0000 & 64 & 12 & 	0.1016 & 	1.9426 \\
194 &   6.0000 & 64 & 10 & 	0.1016 & 	1.9426 \\
243 &   6.0000 & 64 &   8  & 	0.1016 & 	1.9426 \\
260 &   6.0347 & 68 &   8  & 	0.09502 & 	2.0767 \\
265 &   5.8876 & 52 &   6  & 	0.1243 & 	1.5881 \\
275 &   6.0684 & 72 &   8  & 	0.08974 & 	2.1989 \\
285 &   5.9266 & 56 &   6 & 	0.1154 & 	1.7103 \\
290 &   6.1009 & 76 &   8 & 	0.08502 & 	2.3211 \\
305 &   6.1326 & 80 &   8 & 	0.08077 & 	2.4432 \\
324 &   6.0000 & 64 &   6 & 	0.1016	 &      1.9426 \\
366 &   6.0684 & 72 &   6 & 	0.08974	 &      2.1989 \\
397 &   5.8876 & 52 &   4 & 	0.1243	 &      1.5881 \\
428 &   5.9266 & 56 &   4 & 	0.1154	 &      1.7103 \\
458 &   5.9640 & 60 &   4 & 	0.1077	 &      1.8324 \\
486 &   6.0000 & 64 & 	4 & 	0.1016	 &      1.9426 \\
\hline
\end{tabular}
\end{center}
\caption{Lattice setup used for the computation of the
gluon propagator at finite temperature. The $\beta$ was
adjusted to have $ Ls \, a \simeq 6.5$ fm. }
\label{tab1}
\end{table}

In the ultraviolet (UV), we find the propagator is massless and similar to the 1-loop predictions.

In the IR the propagator is compatible with a massive denominator and the simplest fit is a 
Yukawa 
\cite{Oliveira:2010xc}
up to $p \approx 600$ MeV
\begin{equation}
 M_g = 648(7) MeV \ ,
\end{equation}
or a rational function with complex conjugate poles 
\begin{equation}
 M_g=626 \pm i \, 362 \mbox{  MeV} \ .
\end{equation}

\begin{figure} \begin{center}
\includegraphics[width=0.45\textwidth]{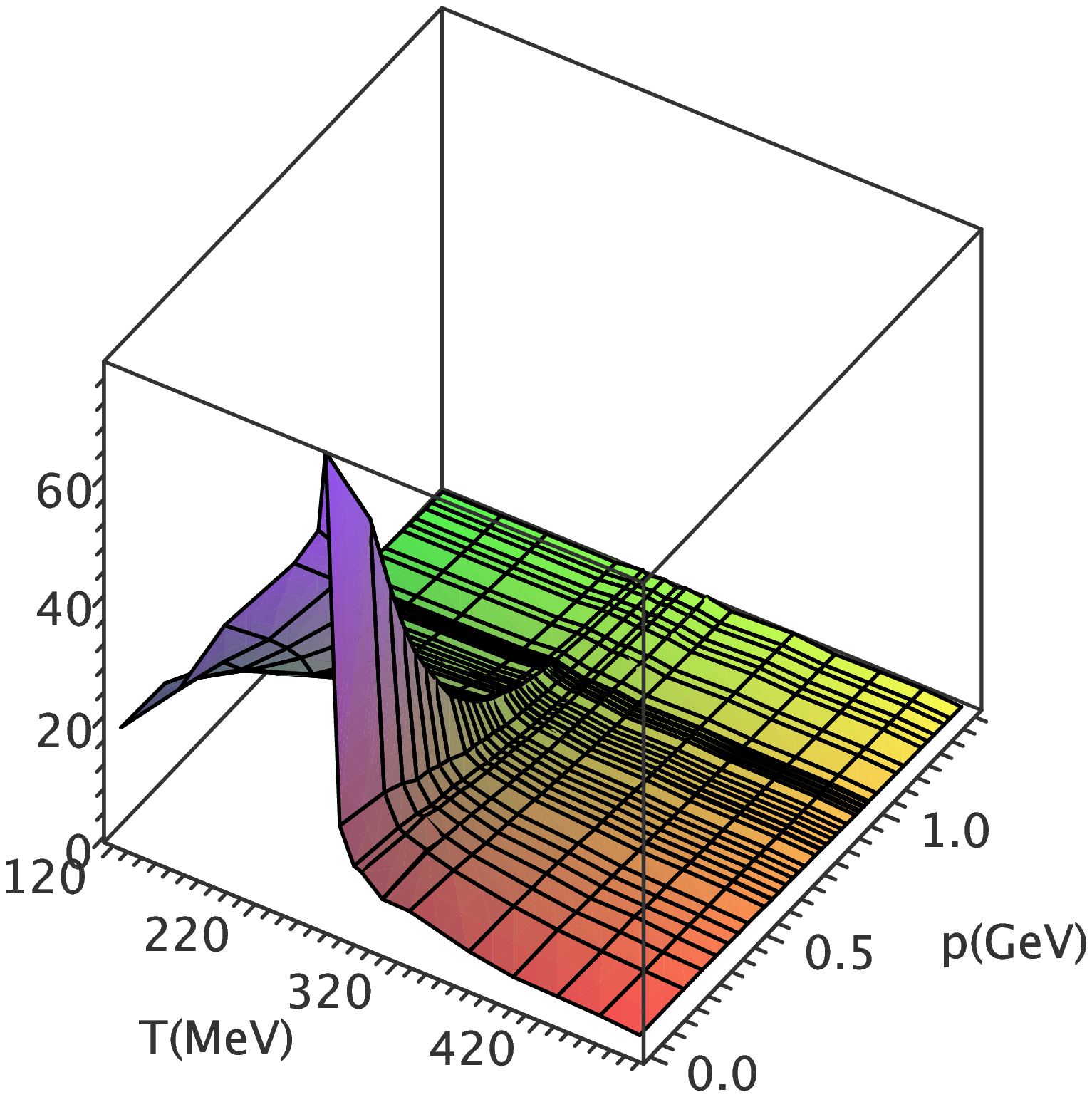}
\includegraphics[width=0.45\textwidth]{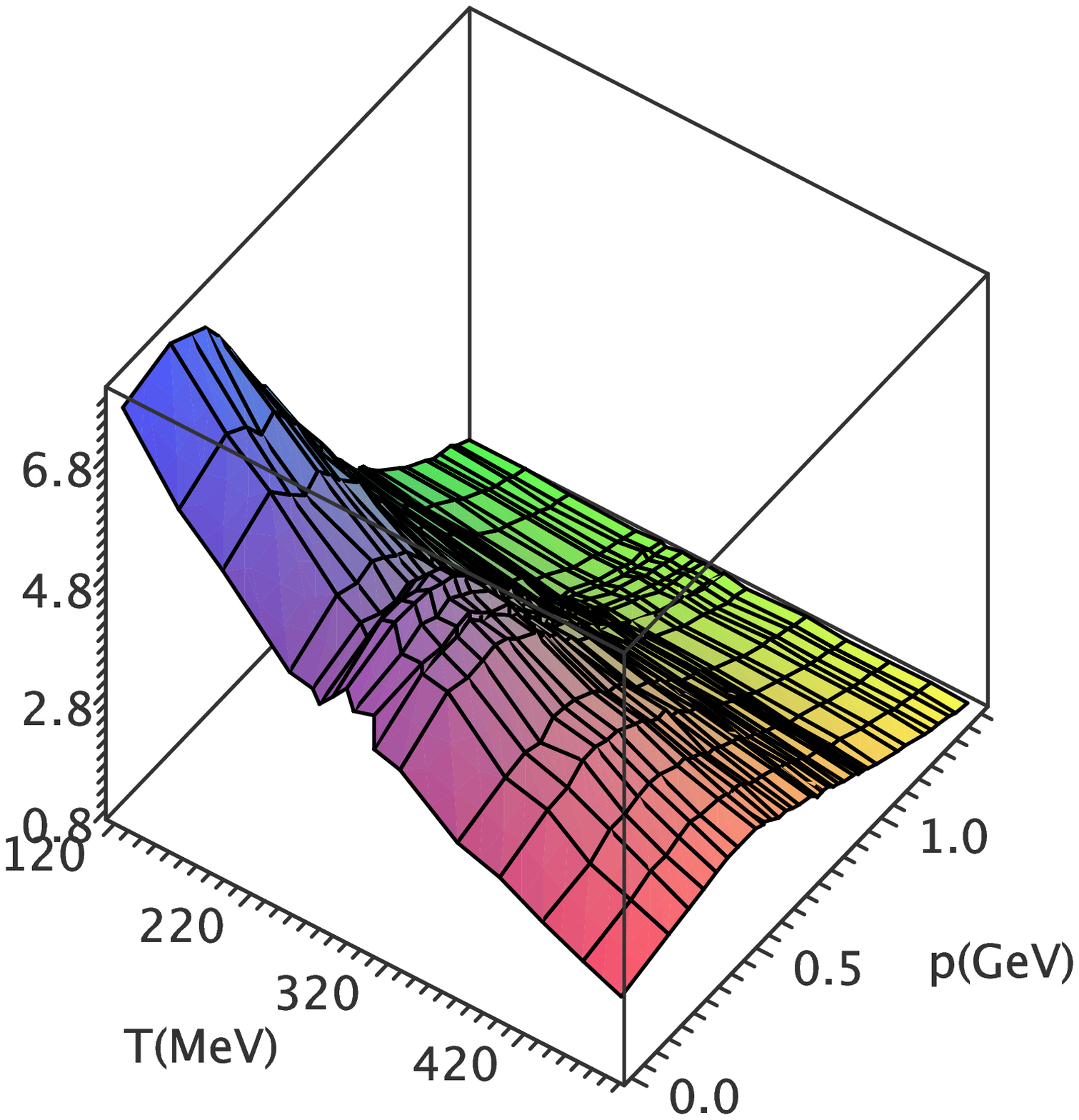}
\end{center} 
\caption{3D plots momenta and temperatures of the longitudinal and transverse propagators.}
\label{fig2}
\end{figure} 

Moreover we also apply a the more elaborate fit of a running gluon mass,
\cite{Oliveira:2010xc}
\begin{equation}
D(p^2) = \frac{ Z(p^2)}{p^2 + M^2(p^2)} \ .
\end{equation}
The running gluon mass is fitted with a parameter 
$m_0 = 723(11)$ MeV 
\begin{equation}
  M^2(p^2) = \frac{m^4_0}{p^2 + m^2_0} \ , \ \ \
Z(p^2) = \frac{z_0}{\left[\log \frac{p^2 + \, r \, m^2_0}{\Lambda^2}\right]^\gamma} \ ,
\end{equation}
and it works up to $p=4.1$ GeV.  This ansatz has a similar functional form to the decoupling solution of the Dyson-Schwinger equations, and to the prediction of the Refined-Zwanziger action 
\cite{Dudal:2010tf}.

\section{Gluon propagator at $T  > 0$} 

At finite $T$,  we project the Lorentz structure of  the propagator $
D^{ab}_{\mu\nu}(\hat{q})$ with two independent form factors,
\begin{equation}
D^{ab}_{\mu\nu}(\hat{q})=\delta^{ab}\left(
P^{T}_{\mu\nu} D_{T}(q_4^2,\vec{q})+P^{L}_{\mu\nu} D_{L}(q_4^2,\vec{q}) \right) \nonumber
\label{tens-struct}
\end{equation}
using transverse and longitudinal projectors in the Landau gauge
\cite{Maas:2011ez,Aouane:2011fv}, 
similar to magnetic and electric projectors respectively, 
\begin{equation}
P^{T}_{\mu\nu} = (1-\delta_{\mu 4})(1-\delta_{\nu 4})\left(\delta_{\mu \nu}-\frac{q_\mu q_\nu}{\vec{q}^2}\right) \nonumber
\label{trans-proj}
\end{equation}
\begin{equation}
P^{L}_{\mu\nu} = \left(\delta_{\mu \nu}-\frac{q_\mu q_\nu}{{q}^2}\right) - P^{T}_{\mu\nu} \nonumber
\label{long-proj}
\end{equation}

\begin{figure} \begin{center}
\includegraphics[width=0.45\textwidth]{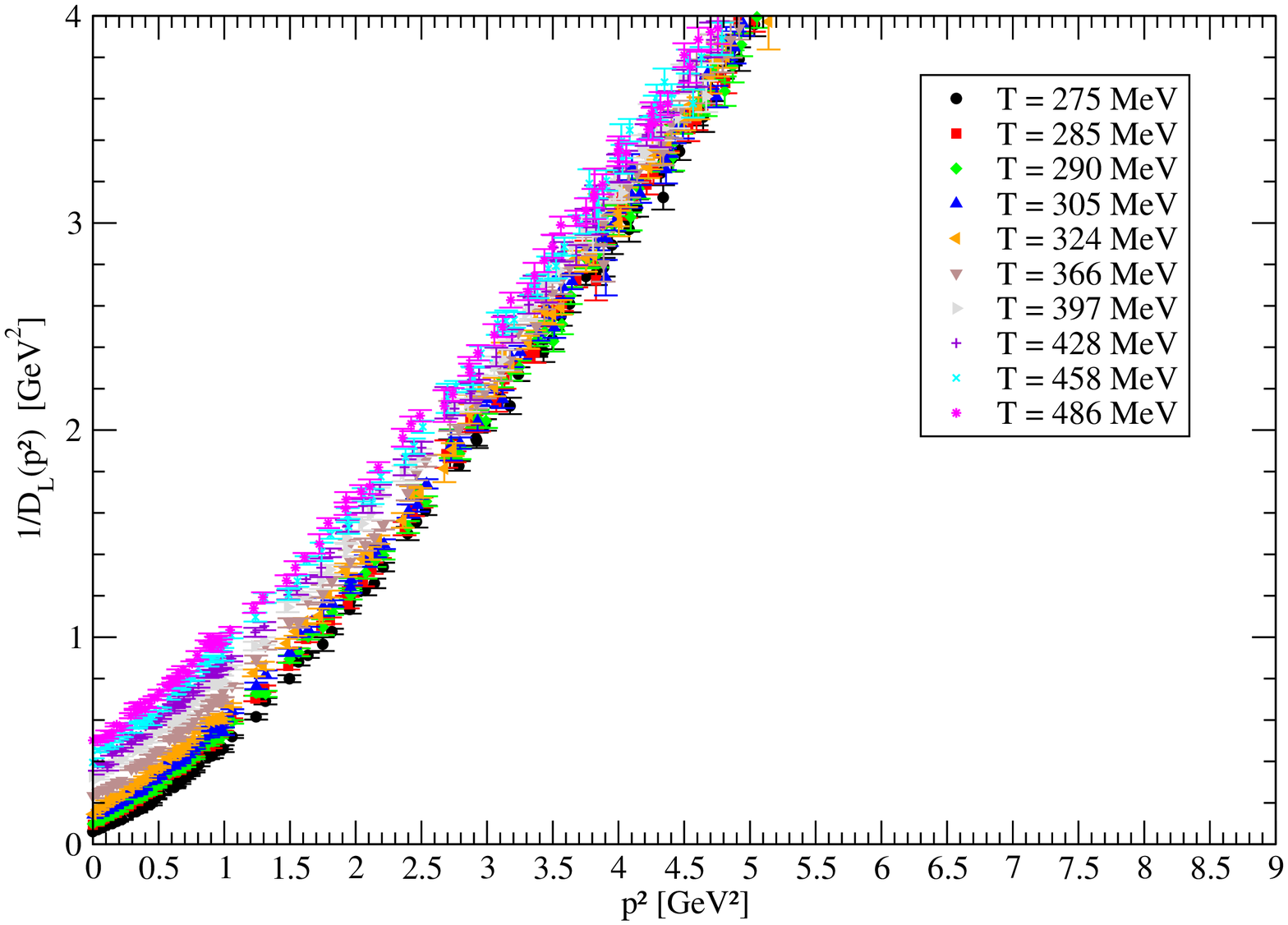}
\includegraphics[width=0.45\textwidth]{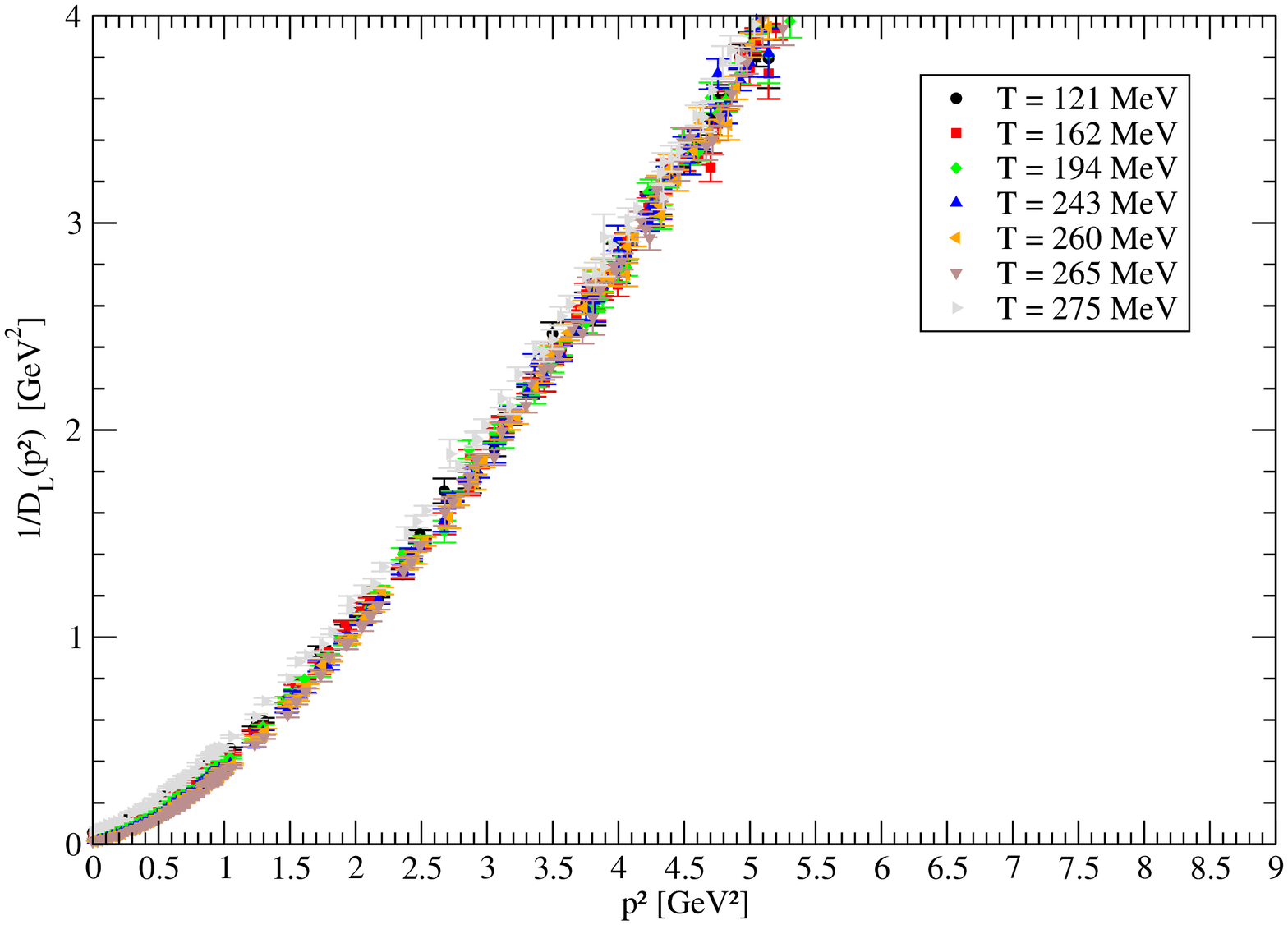}
\includegraphics[width=0.45\textwidth]{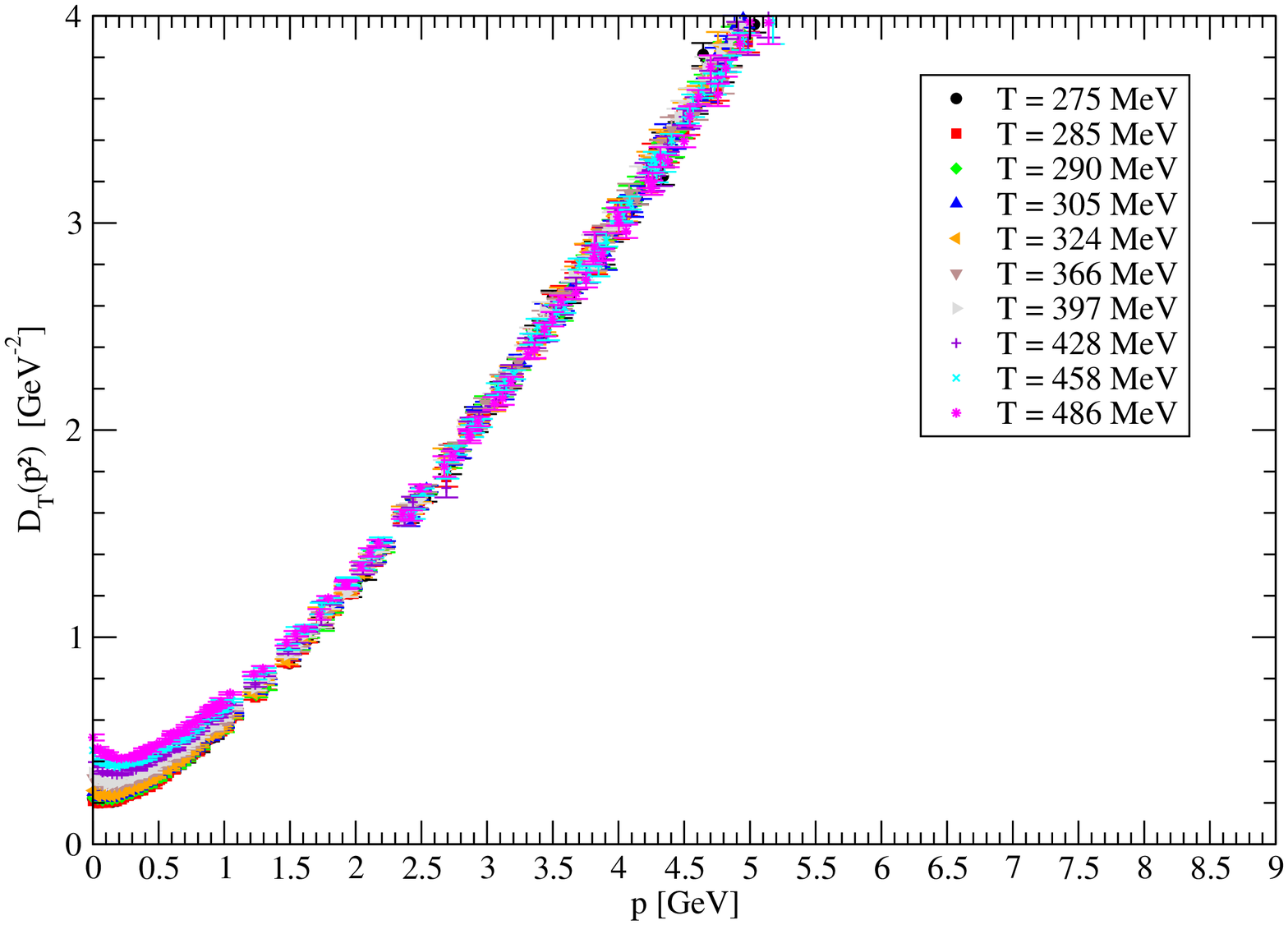}
\includegraphics[width=0.45\textwidth]{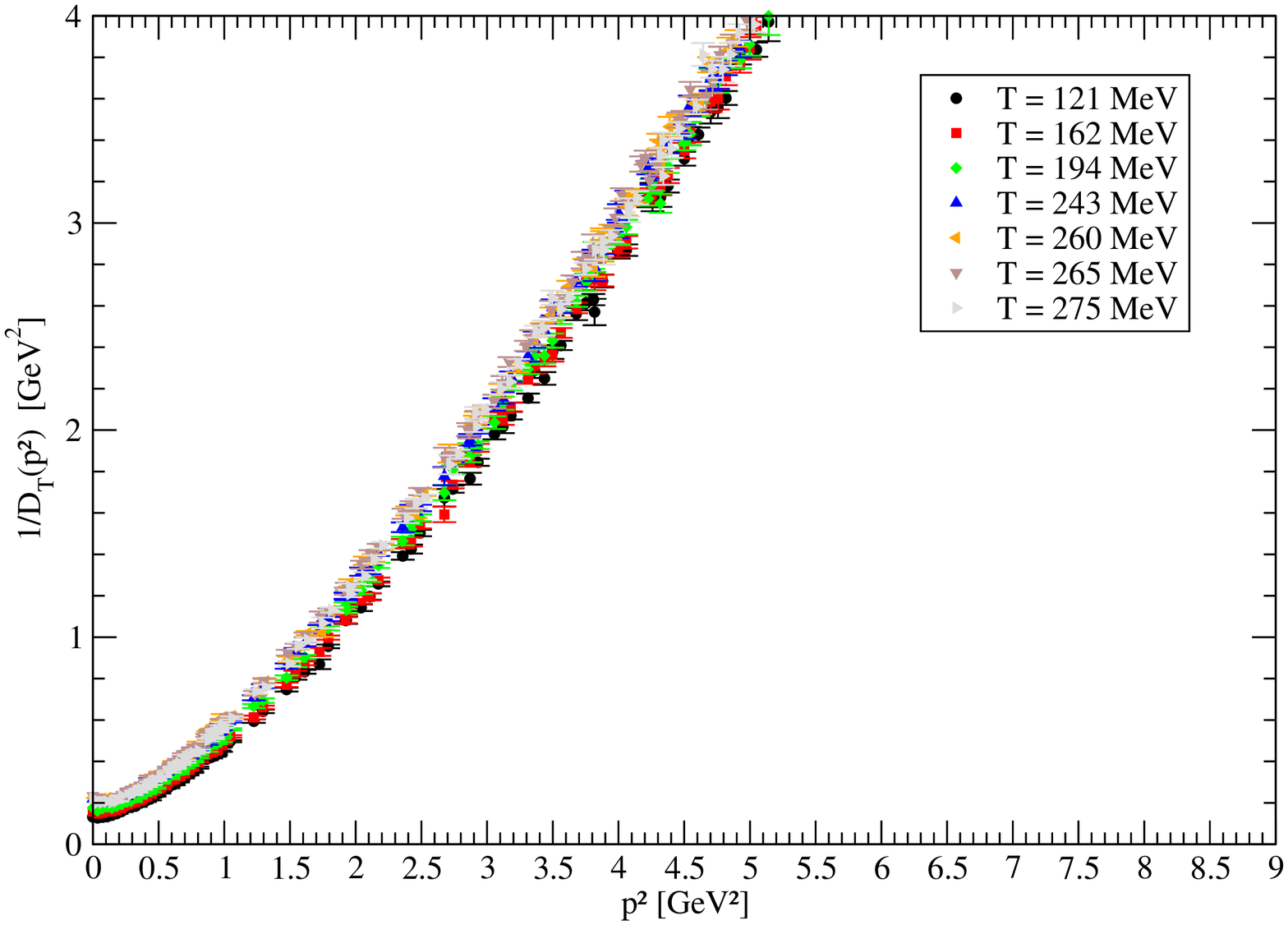}
\end{center} 
\caption{Inverse of the longitudinal and transverse propagators at different finite  temperatures $T$. }
\label{fig3}
\end{figure}

Finite temperature $ T=\frac{1}{a L_t }$ is simply introduced by reducing the extent of temporal direction $ L_t << L_s$. Moreover all lattice data is renormalized fitting the momenta  in the UV region   to the 1-loop inspired propagator,
\begin{equation}
  D_{lattice}(q^2) = \frac{K}{q^2} \left( \ln \frac{q^2}{\Lambda^2} \right)^{-13/22} \ .
\end{equation}
We set  $D(q^2) = Z_R D_{lattice}(q^2)$  where  $D(\mu^2)=1/\mu^2$ with $\mu=$4GeV, in order to 
remove the lattice spacing effects. $D_T$ and $D_L$ are renormalised independently 
but  we observe $Z_L$ and $Z_T$ differ by less than 2 \% .
We utilize the same large volume  with $L_s \sim$ 6.5 fm for all $T$.
Our configurations were generated at the Milipeia and Centaurus clusters of Coimbra University
(Chroma and PFFT libraries).

\section{ Gluon mass at finite $T$}

We plot the finite $T$ inverse of the propagators in Fig. \ref{fig3}.
For IR momenta, they are again compatible with a massive denominator.  
Notice  $D_L^{-1} $ is linear in the infrared, while ${D_T}^{-1}$ bends.
In the UV the propagators have a logarithmica behavior, ${D_i}^{-1} \sim$ log.

\begin{figure} \begin{center}
\includegraphics[width=0.6\textwidth]{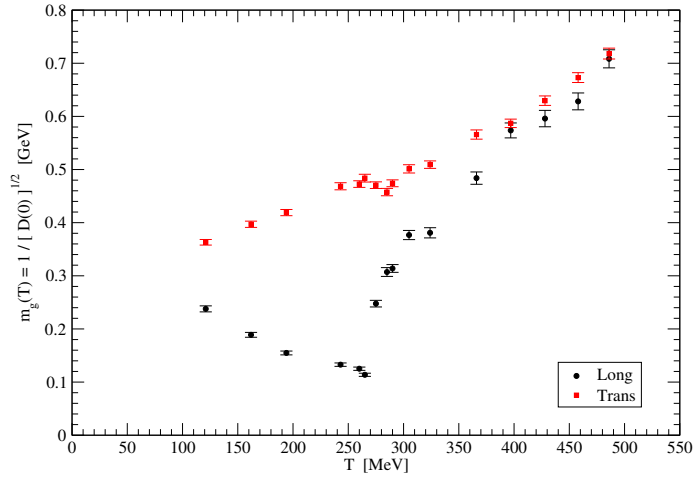}
\end{center}
\caption{Longitudinal and transverse masses fitted with the simplest ansatz at finite temperatures $T$.}
\label{fig4}
\end{figure}

The simplest ansatz for a massive propagator is,
\begin{equation}
 D(p) = {1 \over p^2 + M^2} \ , \ \ 
 \Rightarrow \ \ M = 1 / \sqrt{ D(0) } \ .
\end{equation}
The ersulting fit is shown in Fig. \ref{fig4}. Close to $T_c$,  $D_L$ clearly signals the transition, while $D_T$ is aparently flat.
At $T \sim 2 T_c$, the two masses cross, $M_L \sim M_T$

\begin{table}
\begin{center}
\begin{tabular}{cccccc}
\hline
      $T$        &  $p_{max}$  & $Z_L$  & $M_L$  & $\chi^2/d.o.f.$ \\
      \hline
  121              & 0.467          & 4.28(16)    & 0.468(13)                                   &   1.91 \\
 162              & 0.570          &  4.252(89) & 0.3695(73)                                 &   1.66 \\
 194              & 0.330          &  5.84(50)   & 0.381(22)                                   &   0.72 \\
 243              &  0.330          &  8.07(67)  & 0.374(21)                                   &   0.27 \\
 260              &  0.271          &  8.73(86)  & 0.371(25)                                   &   0.03 \\
 265              &   0.332          &   7.34(45)  &  0.301(14)  &    1.03  \\
 275              &  0.635          &  3.294(65) &  0.4386(83)                               &   1.64  \\
 285              &  0.542          &  3.12(12)   & 0.548(16)                                  &   0.76  \\
 290              &  0.690          &  2.705(50) & 0.5095(85)                                &   1.40  \\
 305              &  0.606          &  2.737(80) & 0.5900(32)                                &   1.30 \\
 324              &  0.870          &  2.168(24) & 0.5656(63)                                &   1.36 \\
 366              &  0.716          &  2.242(55) & 0.708(13)                                  &   1.80 \\
 397              &  0.896          &  2.058(34) & 0.795(11)                                   &  1.03 \\
 428              &  1.112          &  1.927(24)  & 0.8220(89)                                &  1.30 \\
 458              & 0.935          &  1.967(37)  & 0.905(13)                                   & 1.45  \\
 486              & 1.214          &  1.847(24)  & 0.9285(97)                                 & 1.55  \\\hline
\end{tabular}
\end{center}
\caption{Mass $M_L$ and factor $Z_L$ parameters of the Yukawa fits to the longitudinal propagators at finite $T$. }
\label{tab2}
\end{table} 

\begin{figure} \begin{center}
\includegraphics[width=0.6\textwidth]{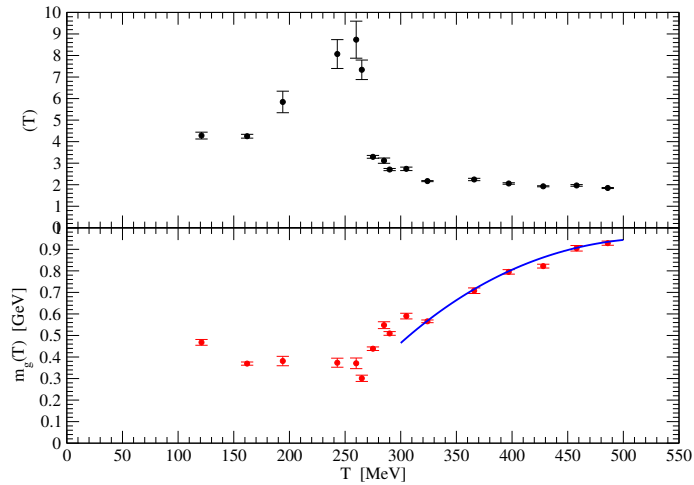}
\end{center} 
\caption{Mass and factor parameters fitted with the Yukawa ansatz at finite temperatures $T$.}
\label{fig5}
\end{figure}

Moreover we apply a better ansatz, adequate for IR momenta, we fit $ D_i$  to a Yukawa with mass $ M$ and dressing function $ Z$
\begin{equation}
  D_i(p^2) = \frac{Z}{p^2 + m^2}
\end{equation}
and look for the largest fitting range $ p_{max}$.
While this fits quite well $D_L$, the Yukawa ansatz does not fit $D_T$. In Fig. \ref{fig5} we show the fit of the mass $m$ and of the factor $Z$. While $Z$ peaks at the transition, the mass $m$ is minimum but clearly finite.

\section{Conclusion}

We compute the gluon propagator in Landau gauge Lattice QCD at finite $ 0 < T <2 T_c$.
The longitudinal component $D_L$ is peaked at $ T = T_c$ .
In the infrared, we fit $D_i$ with massive Yukawa ansatze, the fit to $D_L$ is more stable than the fit to $D_T$.
The fitted longitudinal gluon mass ${M}_L$ is compatible with confinement screening at $T\sim 0$.
${M}_L$  is also consistent with debye screening at $T >> 0$,
We observe $ {M}_L$ is minimum at $T\sim T_c$,  but finite 
\cite{Silva:2013maa}
as suggested by multiplicites  of $\pi$ and $k$ production in heavy ion collisions.


\end{document}